\documentclass[conference]{IEEEtran}
\IEEEoverridecommandlockouts
\usepackage{cite}
\usepackage{amsmath,amssymb,amsfonts}
\usepackage{algorithmic}
\usepackage{graphicx}
\usepackage{textcomp}
\usepackage{xcolor}
\def\BibTeX{{\rm B\kern-.05em{\sc i\kern-.025em b}\kern-.08em
    T\kern-.1667em\lower.7ex\hbox{E}\kern-.125emX}}

\usepackage{xcolor}
\usepackage{balance}
\newcommand{\rd}[1]{\textcolor{black}{#1}}
\newcommand{\bl}[1]{\textcolor{black}{#1}}
\newcommand{\red}[1]{\textcolor{black}{#1}}
\newcommand{\blue}[1]{\textcolor{black}{#1}}
\newcommand{\br}[1]{\textcolor{black}{#1}}
\newcommand{\ma}[1]{\textcolor{black}{#1}}
\newcommand{\df}{{\textsc{WS-Mono3D}}}
\newcommand{\tech}{{\textsc{Mono3D}}}
\newcommand{\conv}{{\textit{Conv}}}

\newcommand{\resc}{{\textit{ResNet-50}}}

\newcommand{\mnet}{{\textit{MobileNetV2}}}

\usepackage{graphicx}
\usepackage[caption=false]{subfig}
\balance

\begin{document}

\title{A New Dataflow Implementation to Improve Energy
Efficiency of Monolithic 3D Systolic Arrays}

\author{
    \IEEEauthorblockN{
        Prachi Shukla\IEEEauthorrefmark{1}
        Vasilis F. Pavlidis\IEEEauthorrefmark{2},
        Emre Salman\IEEEauthorrefmark{3},
        and Ayse K. Coskun\IEEEauthorrefmark{1}
    }
    \IEEEauthorblockA{\IEEEauthorrefmark{1} Boston University - (prachis, acoskun)@bu.edu}
    \IEEEauthorblockA{\IEEEauthorrefmark{2} University of Manchester - vasileios.pavlidis@manchester.ac.uk}
    \IEEEauthorblockA{\IEEEauthorrefmark{3} Stony Brook University - emre.salman@stonybrook.edu}
}

\maketitle

\begin{abstract}
Systolic arrays are popular for executing deep neural networks (DNNs) at the edge. Low latency and energy efficiency are key requirements in edge devices such as drones and autonomous vehicles. 
Monolithic 3D (\tech{}) is an emerging 3D integration technique that offers ultra-high bandwidth among processing and memory elements with \br{a} negligible area overhead. Such high bandwidth can help meet the ever-growing latency and energy efficiency demands for DNNs. This paper presents a novel \ma{implementation} for weight stationary (WS) dataflow in \tech{} systolic arrays, called \df{}. \red{\df{} utilizes multiple resistive RAM layers and SRAM with high-density vertical interconnects to multicast inputs and perform high-bandwidth weight pre-loading while maintaining the same order of multiply-and-accumulate operations as in native WS \rd{dataflow}. Consequently, \df{} eliminates input and weight forwarding cycles and, thus,} provides up to 40\% improvement in energy-delay-product (EDP) over the native WS implementation in 2D at iso-configuration. 
\red{\df{} also provides 10$\times$ improvement in inference per second per watt per footprint due to \blue{multiple} vertical tiers.}
Finally, we also show that \blue{temperature} impacts the energy efficiency benefits in \df{}. 
\end{abstract}


\section{Introduction}
\label{sec:intro}
\begin{figure}[hbt!]
	\centering
\includegraphics[trim= 0mm 0cm 0mm 0cm,clip,width=0.15\textwidth,keepaspectratio]{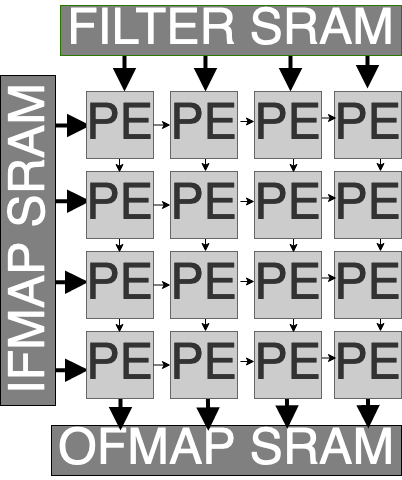}
	\caption{A systolic array: 4$\times$4 PE array with on-chip SRAMs.}
	\label{fig:systolic_array}
\end{figure}

Deep neural networks (DNNs) at the edge have two key goals: latency and energy efficiency. \ma{There are two primary ways} to achieve these goals: (i) increase compute efficiency and (ii) minimize data movement, especially off-chip. Since edge devices are constrained with respect to footprint and \rd{compute/memory} resources, achieving these goals is challenging.
Systolic arrays are among \red{the most} popular DNN accelerator architectures for inference at the edge (Figure \ref{fig:systolic_array}). 
\rd{Systolic arrays are also characterized by a dataflow that defines how the IFMAP, filter weights, and OFMAP are mapped onto the systolic array to minimize data movement and maximize data reuse.}
Weight stationary (WS) is a commonly adopted dataflow
in systolic
arrays 
in which weights are 
first pre-loaded into the processing element (PE) array, 
\cite{jouppi2017datacenter,kung2019systolic},
followed by forwarding of input feature map (IFMAP) to generate the output feature map (OFMAP).
\br{Recently, resistive RAM (RRAM) has gained popularity for storing weights on-chip to eliminate the expensive off-chip DRAM accesses \cite{li2019chip} because RRAM is a high-density CMOS-compatible non-volatile memory (NVM) with low read latency/energy.}

\ma{Monolithic} 3D (\tech{}), an emerging 3D integration technique, has the potential to improve latency and energy efficiency for a variety of DNNs \cite{shukla2021temperature,joseph2021architecture}. \ma{In \tech{} technology}, multiple thin device layers are fabricated sequentially\ma{,} separated by a thin inter-layer dielectric (ILD)
and connected using ultra-thin vertical interconnects, 
called 
monolithic inter-tier
vias (MIV), 
overall providing 
high 
integration density. 
\br{CMOS compatibility in RRAMs also enables a \tech{} integration, further leading to high-bandwidth and high-density edge device\ma{s} \cite{li2019chip}. Furthermore, as an NVM, we can power off some tiers of the system, which lowers the required power and alleviates thermal issues due to vertical integration\rd{, without losing data.}}

We use MIVs for a high bandwidth interface between systolic arrays and on-chip memory \ma{for a new} WS implementation to improve inference latency. We utilize multiple layers \br{of} RRAM to store all weights on-chip and eliminate expensive off-chip DRAM accesses. \br{We also present} design and architectural changes \ma{that result in a novel} WS implementation in \tech{} and improve inference latency and energy efficiency.

Moreover, \tech{} has a shorter heat flow path due to thin layers and dielectric than other 3D technologies, such as die-stacked 3D using TSVs \cite{shukla2021temperature}. However, \blue{multiple layers within \tech{}} can escalate thermal concerns. Also, edge devices may lack a well-equipped cooling system to remove heat from the package. Thus, thermal awareness is key to achieving the energy efficiency promise in \tech{} systolic arrays. 

This work \ma{presents a new WS implementation in \tech{} systolic arrays, \df{}, that is thermally aware and improves latency and energy efficiency.} Prior works on \tech{} systolic arrays \cite{joseph2021architecture,shukla2021temperature} have not considered one or more of the following: (i) the high bandwidth available through MIVs, (ii) high-density RRAM to achieve energy\br{-}efficient DNN acceleration, or (iii) on-/off- chip data movement, which is a significant fraction of system energy. To the best of our knowledge, this is the first work to \br{improve} dataflow implementation by utilizing MIVs for a high-bandwidth interface between monolithically stacked RRAM and systolic arrays to improve latency and energy efficiency. Our contributions are summarized as follows:

\begin{itemize}
    \item We \br{present} \df{}, \ma{a new} WS implementation in \tech{} systolic arrays. \df{} utilizes high-density MIVs to achieve latency and energy efficiency benefits over 2D. It multicasts IFMAP and eliminates 
    IFMAP forwarding. It also enables parallel pre-loading of weights into the PE array, thus eliminating 
    weight forwarding cycles.
    \item We use high-density and high-bandwidth \tech{} RRAM to store all weights on the chip, eliminate DRAM accesses for weights during DNN execution, and enable high bandwidth data transfer between RRAM and PE array using MIVs.
    \item We develop architecture and circuit-level cross-layer models for a 6-tier \tech{} systolic array architecture comprising a PE array, SRAMs for IFMAP and OFMAP, and RRAM layers for storing weights on the chip.
    \item Compared to WS implemented in 2D systolic arrays, \df{} provides up to 47\% and 40\% improvement in latency and energy-delay-product (EDP) for various DNNs for edge applications. The inference per second per watt (I/S/W), inference per second per watt per mm$^2$ (I/S/W/mm$^2$), and inference per second per watt per footprint (I/S/W/footprint) improve by 81\%, 73\%, and 10$\times$, respectively. We also show the thermal impact in MONO3D systolic arrays. E.g., at a strict thermal budget of 75$^\circ$C, EDP benefits reduce to 29\%.
    We also show that the thermal budget plays a vital role in MONO3D systolic arrays with a strict thermal budget. For instance, at a strict thermal budget of 75$^\circ$C, EDP benefits reduce to 29\%.
\end{itemize}
\rd{The rest of the paper is organized as follows. Section \ref{sec:rel_work} briefly discusses WS dataflow, RRAM, and relevant work. We detail \df{} in Section \ref{sec:methodology} and present its evaluation in Section \ref{sec:evaluation}. Finally we conclude and present a discussion on \df{} in Sections \ref{sec:conclusion} and \ref{sec:discussion}, respectively.}

\section{Background and Related Work}
\label{sec:rel_work}
This section presents a background on WS dataflow and RRAM, followed by related work on \tech{} systolic arrays.

\textbf{WS dataflow.}
In WS, weights are first pre-loaded into the PE array \br{from a Filter SRAM through the top edge PEs} \cite{samajdar2020systematic}, then passed to the PE below every cycle, as shown in Figure \ref{fig:systolic_array}). 
After weight pre-loading, IFMAPs are read from the left edge PEs and forwarded to PEs on the right every cycle. Each column in the PE array computes an independent OFMAP channel. PEs generate partial sums (psums) and pass them to the PEs below. The PEs on the bottom edge write outputs back to the OFMAP SRAM. Note that the outputs from different columns belong to different OFMAP channels. Often, there is an insufficient number of PEs to map the whole compute. In such cases, computation is sliced into folds ($F$)
\cite{samajdar2020systematic}. Consequently, the compute cycles in WS can be broken down as shown in Eq. \eqref{eq:ws}. 
\begin{equation}
\footnotesize
    \label{eq:ws}
    C_{WS} = \sum_{i} (w_i + I_i + O_i),
\end{equation}
where $C_{WS}$ is the compute cycles in WS, $1 \leq i \leq F$ folds, $w_i$ is the number of cycles spent in pre-loading weights, and $I_i$ is the number of cycles to forward IFMAP from left to right until all of the pre-loaded PEs have IFMAP to generate psums. $O_i$ includes compute cycles when all the pre-loaded PEs are generating psums (i.e., maximum throughput) and cycles spent forwarding psums from top to bottom.

\rd{\textbf{Resistive RAM.}}
\rd{RRAM is a high-density CMOS-compatible} 
\rd{emerging non-volatile memory with low read latency/energy 
but has write endurance issues. }
\rd{Due to these characteristics, RRAMs are also getting popular in edge DNN accelerators 
for storing 
\begin{figure}[hbt!]
\subfloat[\df{} stack.]{
\label{fig:stack}
\includegraphics[trim= 0mm 0cm 0mm 0cm,clip,width=0.27\textwidth,keepaspectratio]{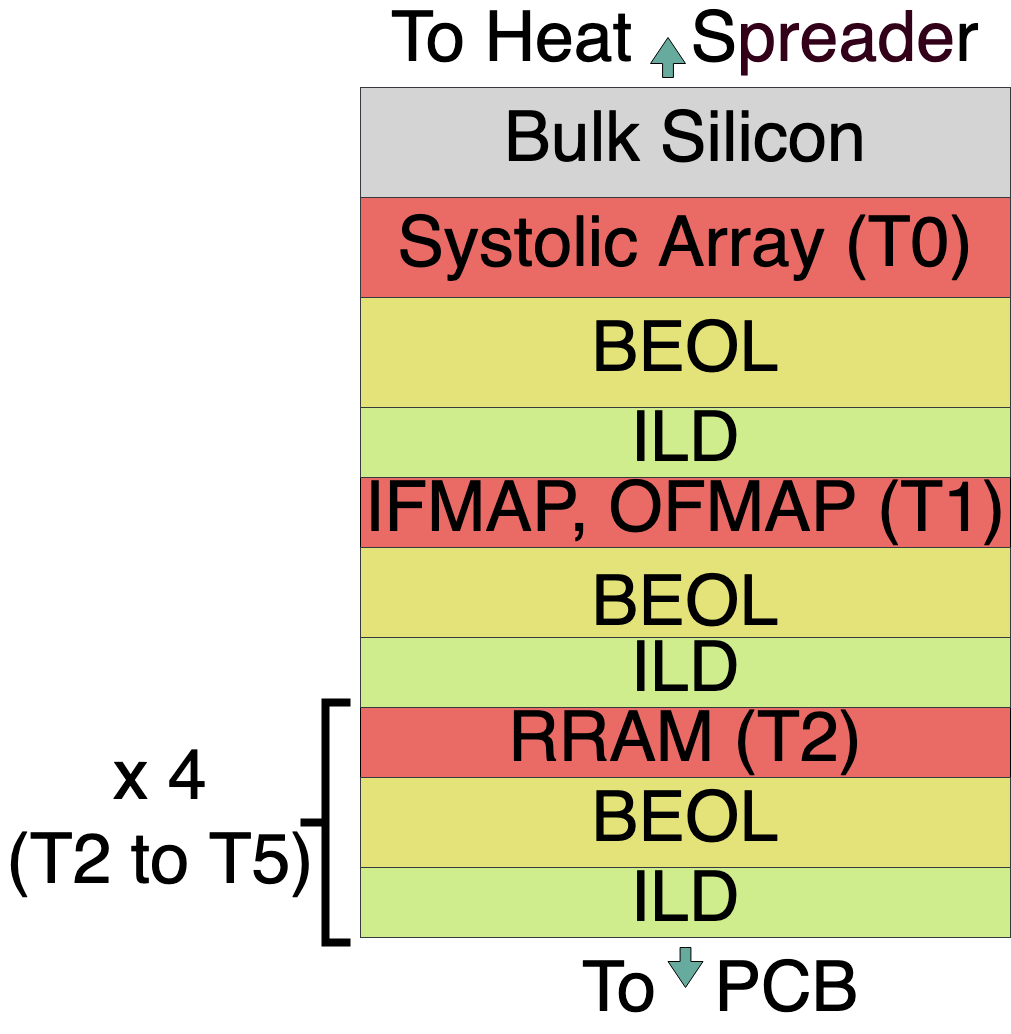}
}
\subfloat[Top view: 2D floorplan.]{
\label{fig:2d_flp}
\quad\includegraphics[trim= 0mm 0cm 0mm 0cm,clip,width=0.14\textwidth,keepaspectratio]{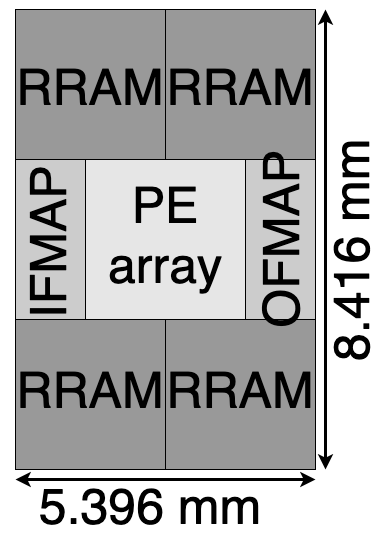}\hfill
}
	\caption{(a) A flip-chip 6-tier \tech{} chip stack with 4 RRAM tiers for storing weights. Each tier is 2.816$\times$2.816 mm$^2$, (b) Top view of (a)'s 2D counterpart.}
	\label{fig:array_stack}
\end{figure}
weights on-chip \cite{aly2018n3xt,li2019chip}. 
A high-resistance state in an RRAM cell encodes bit `0', while a low resistance state encodes bit `1'. An RRAM cell can also encode multiple bits per cell\bl{, i.e., multi-level cell (MLC). However, endurance issues are more pronounced in MLC devices, and hence, are not considered in this work \cite{lee2012multi}.}
Furthermore, RRAM can be fabricated with \tech{} technology \cite{aly2018n3xt}. 
\bl{In this work, we model a large-capacity multi-layer RRAM to eliminate off-chip DRAM accesses during a DNN execution by storing weights on-chip.}}

\textbf{Related Work.} \ma{Existing research effort in} \tech{} DNN accelerators focuses on important aspects of \tech{} accelerator design, e.g., weight/activation sparsity, SRAM partition choices, process variation, compute-in-memory \cite{murali2020heterogeneous,shukla2021temperature,yu2020spring}. However, none of them has exploited the ultra-high bandwidth in \tech{} technology to improve dataflows for more efficiency.
The closest work by Joseph et al. \cite{joseph2021architecture} distributes output stationary (OS) dataflow in 3D systolic arrays. It divides the PE array across eight tiers and assigns private SRAMs to each tier. However, this approach leads to duplicate IFMAP storage across tiers, which they do not address. Its area, energy, and performance models include only the PE array without considering the on-chip SRAMs, DRAM, or interconnects, thus making the evaluation incomplete. 

\br{Both OS and WS are commonly used and have different tradeoffs. E.g., while OS provides lower latency, it also has a high bandwidth requirement to support a stall-free execution \cite{samajdar2020systematic}. On the other hand, WS has a higher latency but requires lower bandwidth, and also results in higher utilization of systolic array \cite{samajdar2020systematic}. Hence, it \ma{can be} misleading to determine a winner among them, especially due to the unexplored traits of 3D technologies, which we \ma{analyze} and exploit here.}
In this paper, we optimize WS for \tech{} systolic arrays and evaluate its benefits over 2D using detailed cross-layer architecture- and circuit-level models.

\begin{figure}[hbt!]
    \centering
    \includegraphics[trim= 0mm 0cm 0mm 0cm,clip,width=0.48\textwidth,keepaspectratio]{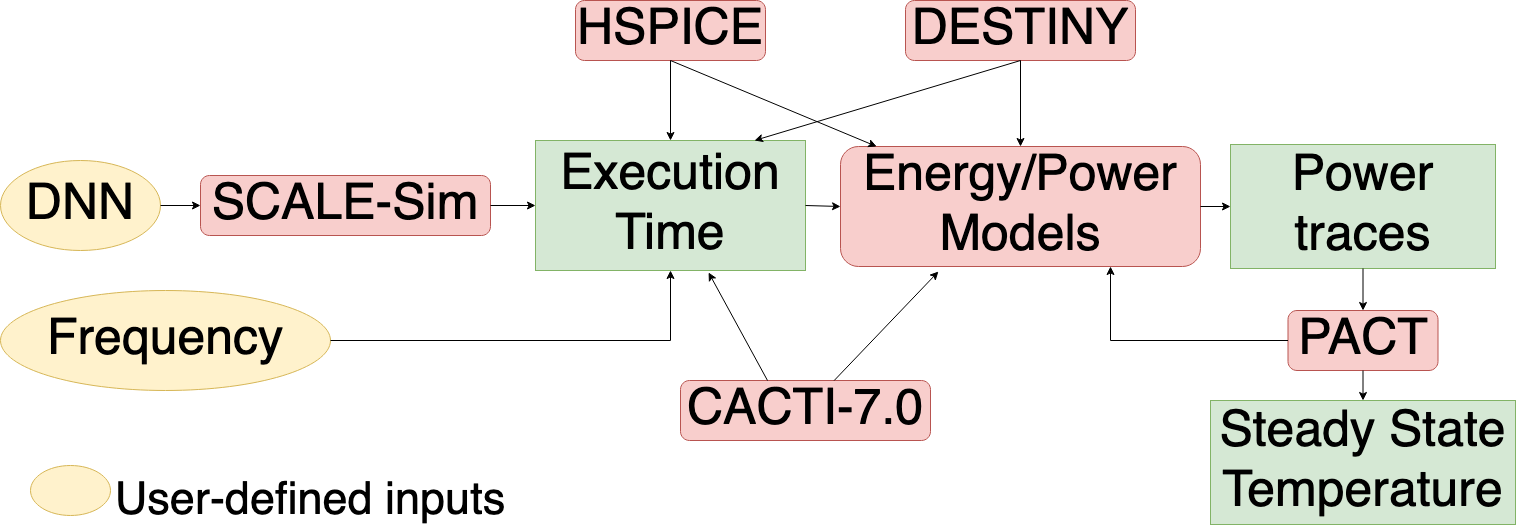}
    \caption{Evaluation framework for \df{}}
    \label{fig:framework}
\end{figure}
\section{\df{}}
\label{sec:methodology}

We begin with an overview of the DNNs investigated in this paper and the \tech{} chip stack. We then \br{describe} our 
\ma{improvements} in \df{}. Finally, we \br{detail} our cross-layer architecture- and circuit-level models for performance, power, temperature, and area to evaluate \df{}.

\subsection{Overview}
\label{sec:overview}

To \ma{evaluate \df{} benefits}, we target six high-accuracy DNNs commonly deployed at the edge: ResNet-18, ResNet-32, ResNet-50, MobiLeNet-V1, EfficientNet-B0, and GoogLeNet. Since the topologies of these DNNs vary from one another, their execution leads to varying performance, power, and thermal profiles.
Figure \ref{fig:stack} shows the \blue{flip-chip} 6-tier \ma{\tech{} stack,} we investigate in the paper. 
To demonstrate the benefits of \df{}, we choose a 256$\times$256 systolic array with 2 MB IFMAP SRAM, 2MB OFMAP SRAM, and 32 MB Filter RRAM as our test vehicle. We select this configuration with the objective to (i) have sufficient on-chip memory capacity to \br{eliminate} off-chip DRAM accesses during DNN execution, (ii) minimize RRAM endurance concerns by including sufficient capacity to store all weights without overwriting any cell during DNN execution, and (iii) minimize area mismatch between tiers (including MIV overhead). Tier 0 with the systolic array is closest to the heat spreader because it has highest power consumption \br{($Pc$)} among all tiers. 
\br{Although a 6-tier \tech{} system is challenging from a manufacturing perspective, specific and strong arguments support its exploration. First, \tech{} offers ultra-dense integration. Since 2D technology is approaching its scaling limits, \tech{} is a potent technology for designing DNN edge devices with area and bandwidth constraints \cite{zhang2019recent,lee2020hardware}. Second, encouraging demonstrations of monolithic integration of SRAM, logic, and RRAM have been shown \cite{andrieu2018review,srimani2020heterogeneous}. In this work, we assume a mature \tech{} technology where a 6-tier stack will be possible. Third, recent works have explored multiple-layer Mono3D architecture\ma{s} in designing caches, DNN accelerators to demonstrate the potential of this technology with respect to latency, bandwidth, and integration density \cite{yu2020spring,joseph2021architecture}.}

\subsection{\ma{\df{} Implementation Decisions}}
We make three main architectural \ma{design decisions} to utilize MIVs and \ma{improve the} spatio-temporal WS characteristics in \tech{}: (i) $A_1$: vertical integration of SRAMs and RRAMs for high-bandwidth interface with the PE array; (ii) $A_2$: In every WS fold, reduce the number of weight preload cycles to one by reading all the weights into the PE array; (iii) $A_3$: In every WS fold, reduce the IFMAP forwarding cycles to one by multicasting the IFMAPs to all the PEs in their respective rows. As a result of these decisions, Eq. \eqref{eq:ws} reduces to $C_{WS-Mono3D} = \sum_{i=1}^{F} (1 + 1 + O_i)$. In $A_1$, we update RRAM bank architecture to eliminate the H\ma{-}tree horizontal routing for data bits from the I/O port to the center of the bank \cite{poremba2015destiny}.
With vertical vias, we assume that the data bits arrive at the center of an RRAM bank rather than going to the port at the edge. Furthermore, since RRAMs have dedicated tiers in our chip stack, we implement the H\ma{-}tree routing in their corresponding BEOL. These RRAM architecture decisions eliminate the area overhead resulting from high bandwidth RRAM.

\subsection{Architecture- and Circuit-level Cross-layer Models}
Figure \ref{fig:framework} shows our cross-layer modeling framework to evaluate \df{}. 
We have architecture-level \br{area}, performance, 
and power models for DNN inference on systolic arrays, SRAMs, and RRAM.
Circuit-level
models comprise delay and power models for MIV, interconnect, and inter-PE communication.

For temperature estimation\ma{,} we use a compact thermal simulator. 
For performance evaluation, we model \df{} in SCALE-Sim \cite{samajdar2020systematic}. SCALE-Sim is a CNN simulator for systolic arrays that models a stall-free inference. It models double-buffered on-chip memory to hide the DRAM cycles during DNN execution.
We generate per-fold counters to determine the weight preloading cycles ($w_i$) and IFMAP forwarding cycles ($I_i$), and then calculate $C_{WS-Mono3D}$ for each DNN.
For every convolutional (\conv{}) layer, in addition 
to compute cycles, SCALE-Sim outputs non-overlapping DRAM cycles that contribute towards total execution cycles.
Since we model sufficient on-chip RRAM/SRAMs to store the inputs and outputs during a \conv{} layer execution, we only add the non-overlapping DRAM cycles of the first $Conv$ layer (read inputs) and the last layer (write outputs) to calculate total execution cycles.
For the two layers, we add additional cycles due to RRAM and SRAM read/write latencies plus routing delay to reach the on-chip memory tiers from tier 5 (i.e., PCB side in Figure \ref{fig:stack}). The routing delay includes delays due to lateral and vertical distances (estimated using Manhattan Distance modeling) and calculated using HSPICE. Finally, we calculate the total DNN execution time using the user-defined frequency.
We use SCALE-Sim's default DRAM bandwidth of 10\,B\,per\,cycle.  

We use DESTINY \cite{poremba2015destiny} and CACTI-7.0 \cite{thoziyoor2009cacti} to \br{model RRAM and SRAMs, respectively, and generate their area, latency, and $Pc$.} Each SRAM is 2 MB with 16 B word length and 16 banks.
We determine RRAM dynamic read/write energy and leakage using DESTINY. Since DESTINY models only a single bank, we assume each RRAM tier comprises \br{64 banks, each of 128 KB capacity and 256 B word length.} Each SRAM/RRAM bank can be accessed in parallel, and has one read and one write port, each with dedicated MIVs. 
\br{We assume one MIV per bit. Since MIVs are nanometer scale and incur minimal area overhead, this assumption is reasonable.}
\br{\#MIVs in each SRAM/RRAM bank equals the size of address and data buses for each port, and the MIV area overhead is added to each bank. E.g., each RRAM bank has 2$\times$2,057 MIVs (9 for address and 2048 for data).}
For interconnect power modeling, \br{inter-}RRAM \br{routing} \br{$Pc$} is assigned to the tier's BEOL.

\br{All address bits arrive at each SRAM/RRAM bank at the edge (default model in DESTINY/CACTI due to peripheral logic). However, due to the ultra-dense RRAM bandwidth, we assume the data bits arrive at the center instead to save read latency and energy. Thus, we update DESTINY’s RRAM model by setting the edge-to-center delay and power to 0 for the data bus. We make a simplifying assumption that the data and address bits first route vertically through the MIVs and then laterally in the systolic array tier. We use Manhattan Distance to calculate wirelengths (a commonly used approach). 
Also, RRAM routing through its metal tiers is an option already provided by DESTINY to reduce area overhead. Due to \ma{page} limitations, we have not added figures.}
Thus, routing $Pc$ due to lateral wirelengths is added to the systolic array BEOL, while the MIV power is added to SRAM/RRAM tiers' BEOL. In addition, we use HSPICE and array utilization to calculate the inter-MAC IFMAP, weight, and OFMAP forwarding $Pc$. \br{While we add all three forwarding $Pc$s to the PE array power in WS in 2D, we add only the OFMAP forwarding $Pc$ in \df{} due to the architectural decisions.} 

\br{We use a floorplan for the aimed 6-tier system where the area numbers are generated by the architecture-level tools.} \ma{Physical design is out of scope in this letter, but there is ongoing research on \tech{} PDKs.} 
Finally, for steady-state temperature estimation, we build a thermal model for our \tech{} system in PACT, our in-house open-source SPICE-based compact thermal simulator \cite{yuan2021pact}. For accurately determining leakage, we run DESTINY/CACTI iteratively with PACT-generated temperatures until \br{convergence}, i.e., the temperature difference between consecutive runs $<$ 1$^\circ$C. \br{In our analysis at 22 nm, a change of 1$^\circ$C has a negligible impact on leakage. Thus, a smaller convergence criterion may be chosen but will not impact the thermal and power estimation results and instead result in longer simulations.}

\section{Evaluation}
\label{sec:evaluation}
This section first describes the experimental setup in \df{}, and then discusses its benefits with respect to 2D WS.
\subsection{Experimental Setup}
\label{sec:setup}

We perform our analysis at \red{22 nm} CMOS technology node \br{to demonstrate \tech{} benefits because of the availability of open-source tools that we utilize in this letter.}
We use a representative MAC unit area, energy, and frequency values from a recent work \cite{shukla2021temperature}: 121 $\mu$m$^2$, 0.26 pJ per 8-bit \br{integer} MAC operation, 1 GHz. Tier 0 is 500 nm in thickness, while the height of upper tiers \blue{is} determined by gate pitch, i.e., 8$\times \frac{technology\ node}{2}$ $\approx$ 85 nm \cite{bhattacharya2016ultra}. The length of an MIV is 270 nm since it passes through the ILD (100 nm), upper tier (85 nm), and the dielectric between the tier and metal layer (85 nm). We also use representative values for MIV's diameter, pitch, area, resistance, and capacitance \cite{samal2016monolithic}. 
Using HSPICE, we obtain (i) MIV delay and energy values of 8.6 ps and 0.02 fJ, (ii) inter-MAC delay and energy values of 14 ps and 0.08 fJ are the delay and energy between neighboring PEs, respectively.

We evaluate \df{} for \bl{DNN inference, using} the six DNNs studied in this paper at three frequency levels: 500 MHz, 700 MHz, and 1000 MHz. \br{While 1 GHz is from a representative recent work on mobile systolic arrays \cite{li2019chip}, the other two frequency levels are chosen to demonstrate \tech{} impact on I/s/W and temperature at different frequency levels, mimicking dynamic frequency scaling mechanisms that are commonplace in modern mobile products.}
\bl{We assume a batch size of 1 for DNN inference \cite{kwon2021heterogeneous}.}
 To compare \df{} to WS, we model a 2D 256$\times$256 systolic array with iso-capacity SRAM and RRAM that implements WS dataflow. A top view of the 2D floorplan is shown in Figure \ref{fig:2d_flp}. 
 All forwarding energies and power within the PE array are added to the 2D WS setup. Chip footprint dimensions in \tech{} setup is 2.816 mm$\times$2.816 mm, while the 2D setup's dimensions are 8.416 mm$\times$5.398 mm. 
To model the absence of heat sinks on edge devices, we reduce its thickness to 1 nm. The heat spreader thickness is set to 1 mm, and 45$^\circ$C is the ambient temperature. We also use two thermal budgets, 75$^\circ$C and 85$^\circ$C, to evaluate the thermal effects on \df{}. 
\br{To model a low-cooling capability, we use a poor convection resistance (1.3 W/$^\circ$C) \cite{skadron2003temperature}.}


\begin{figure*}[t]
\centering
\subfloat[Latency (ms) in \df{}.]{
\label{fig:latency_mono3D}
\includegraphics[trim= 0mm 0mm 0mm 0mm,clip,width=0.46\textwidth,keepaspectratio]{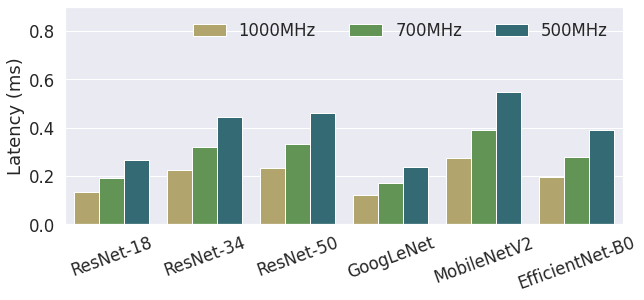}
}
\subfloat[Latency (ms) in WS.]{
\label{fig:latency_2D}
\includegraphics[trim= 0mm 0cm 0mm 0cm,clip,width=0.44\textwidth,keepaspectratio]{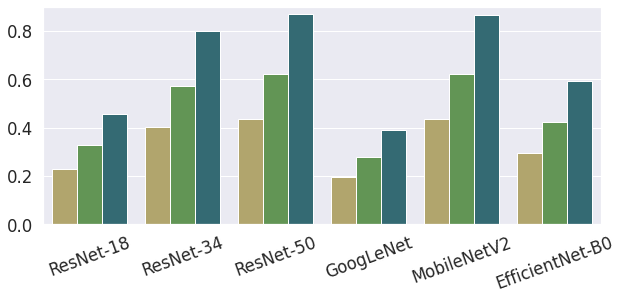}
}

\subfloat[Chip Power (W) in \df{}.]{
\label{fig:power_mono3D}
\includegraphics[trim= 0mm 0cm 0mm 0cm,clip,width=0.46\textwidth,keepaspectratio]{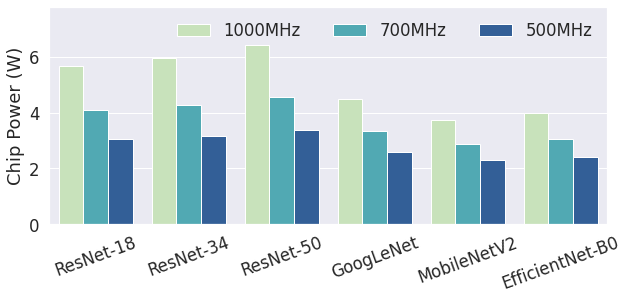}
}
\subfloat[Chip Power (W) in WS.]{
\label{fig:power_2D}
\includegraphics[trim= 0mm 0cm 0mm 0cm,clip,width=0.44\textwidth,keepaspectratio]{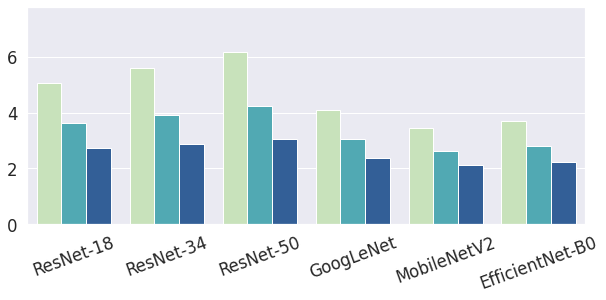}
}

\subfloat[\df{} Steady State Temperatures.]{
\label{fig:temp_mono3D}
\includegraphics[trim= 0mm 0cm 0mm 0cm,clip,width=0.46\textwidth,keepaspectratio]{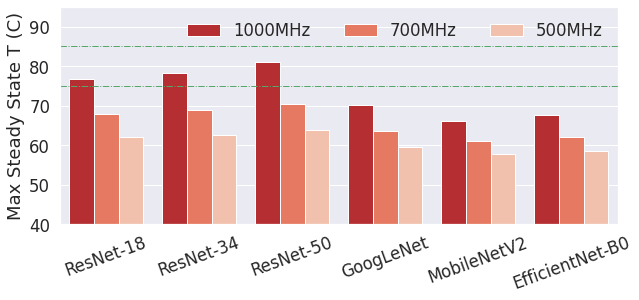}
}
\subfloat[EDP Benefits in \df{} w.r.t. WS.]{
\label{fig:edp_mono3D}
\includegraphics[trim= 0mm 0cm 0mm 0cm,clip,width=0.46\textwidth,keepaspectratio]{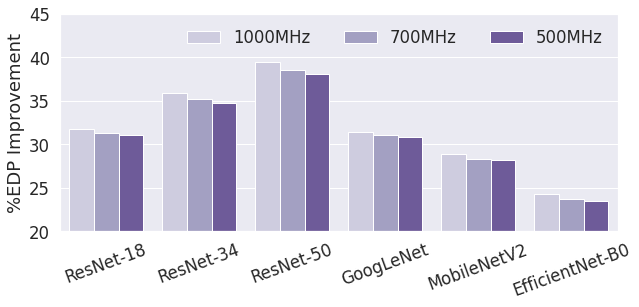}
}

\subfloat[I/s/W/area improvement in \df{} w.r.t. 2D WS.]{
\label{fig:IpsArea_mono3D}
\includegraphics[trim= 0mm 0cm 0mm 0cm,clip,width=0.46\textwidth,keepaspectratio]{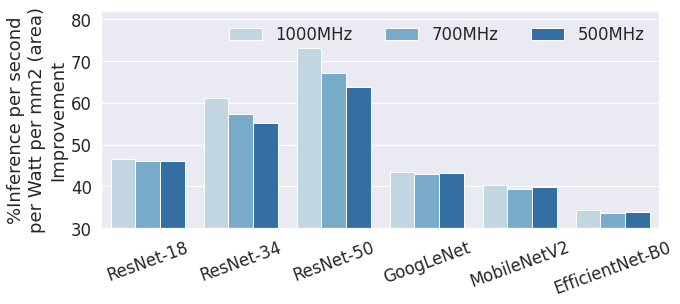}
}
\subfloat[I/s/W/footprint improvement in \df{} w.r.t. 2D WS.]{
\label{fig:IpsFootprint_mono3D}
\includegraphics[trim= 0mm 0cm 0mm 0cm,clip,width=0.46\textwidth,keepaspectratio]{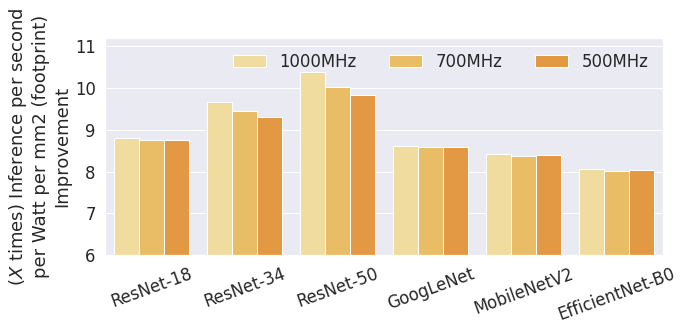}
}
  \caption{\df{} versus WS in 2D for several DNNs at three frequency levels. (a-b) show absolute inference latencies in ms. Latencies in \df{} are up to 47\% lower. (c-d) show absolute power values in Watt (W). (e) shows steady state temperatures in \df{} with dotted lines for \red{two} thermal constraints. (f) Up to 40\% EDP benefits in \df{} w.r.t. WS in 2D. \rd{(g) Up to 73\% improvement in I/p/s/area in \df{} w.r.t. WS in 2D. (h) Up to 10$\times$ improvement in I/p/s/footprint in \df{} w.r.t. WS in 2D.}} 
  \label{fig:results}
\end{figure*} 

\subsection{Results}
\label{sec:results}
We compare \df{} to 2D WS at iso-frequency to evaluate inference latency and energy efficiency benefits. Finally, we also demonstrate that thermal awareness plays an important role in the design of systolic arrays implementing \df{}. 

Figs. \ref{fig:latency_mono3D}-\ref{fig:latency_2D} and \ref{fig:power_mono3D}-\ref{fig:power_2D} show the absolute inference latencies and chip power for the six DNNs, respectively. \br{The total system energy (chip + DRAM) is comparable between \df{} and WS.} \df{} achieves a latency reduction of up to 47\% (avg. 41\%) due to a reduction in compute cycles from IFMAP multicast and parallel weight preloading. \df{} has up to 12\% (avg. 9\%) higher chip power than 2D WS. This is primarily because more RRAM banks are active for the parallel preloading of weights using MIVs. Overall, \df{} achieves up to 40\% reduction in system EDP (avg. 32\%) with respect to WS in 2D, also shown in Figure \ref{fig:edp_mono3D}. Note that system EDP also includes DRAM energy. 
\rd{Interestingly, \df{} benefits with respect to the EDP are greatest in \resc{}. This is due to two reasons. First, \df{} provides more significant benefits in \conv{} layers than fully-connected (FC) layers. FC layers are matrix-vector multiplication, where only the first row in a systolic array is utilized. Consequently, \df{} provides improvement only due to the multicasting of the inputs. In contrast, \conv{} layers are matrix-matrix multiplication and can benefit from both multicasting and parallel pre-loading of weights. Second, \df{} benefits increase with \bl{a greater} number of DNN channels. \bl{Greater number of channels means more cycles are spent in left-to-right input forwarding in 2D WS and, hence, more benefits can be achieved from input multicasting in \df{}.} Since, out of all the DNNs investigated in this paper, \resc{} has the maximum number of \conv{} layers (i.e., 48) with the number of channels ranging from 64 to 2048, \df{} benefits are the highest.}

\df{} also provides improvement up to 81\% (avg. 55\%) in I/s/W over WS, primarily due to the latency benefits in \df{}. For area- and energy- efficiency, we also report 73\% (avg: 48\%) I/s/W/area improvement over WS in 2D, which includes total silicon area, also shown in Figure \ref{fig:IpsArea_mono3D}. I/s/W/mm$^2$ improvements reduce slightly because the \df{} chip stack has $\approx$1 mm$^2$ area overhead due to MIVs.
In addition, we report 10$\times$ (avg: 9$\times$) I/s/W/footprint improvement in Figure \ref{fig:IpsFootprint_mono3D}\br{, out of which the footprint benefit is $\approx$6$\times$. Note that both the \tech{} and 2D footprints are specified in Section \ref{sec:setup}.}
While footprint efficiency is critical due to limited package area, it comes with an additional fabrication cost for the vertical tiers. 
Since we do not model the cost, we present a conservative comparison using the total silicon area, and \bl{leave a detailed cost model as future work.} 

We also obtain steady state temperatures and evaluate \df{} at various thermal constraints. \ma{A} relaxed constraint of 85$^\circ$C allows DNN execution \ma{at} all three frequencies. However, under tighter constraints, e.g., 75$^\circ$C, 
the average latency and EDP benefits reduce to 29\% and 18\%, respectively. This is because while ResNets execute at 1000 MHz in the 2D systolic array with WS dataflow, the strict thermal budget allows 700 MHz (not 1000 MHz) in \df{} to avoid thermal violations.
Thus, \ma{temperature impacts \df{} benefits.}

\section{Conclusions and Future Work
\label{sec:conclusion}}
This paper presents \df{}, a \ma{novel} implementation of WS dataflow in \tech{} systolic arrays. \df{} utilizes the ultra-high \tech{} bandwidth in \tech{} technology and eliminates cycles spent in pre-loading weights and forwarding IFMAPs. To evaluate \df{}, we investigate a 6-tier \tech{} chip stack with 256$\times$256 PE array, 4 MB SRAMs for IFMAP and OFMAP, and 32 MB RRAM for weights, for several edge DNNs. Compared to WS in 2D, 
\df{} provides up to 47\% reduced latency and 40\% lower EDP at a relaxed temperature constraint of 85$^\circ$C. 
However, at a tighter thermal constraint of 75$^\circ$C, these reduce to 29\% and 18\%. This demonstrates a need for thermal awareness in the design of \df{} systolic arrays. 
\blue{We also demonstrate up to 81\% improvement in I/s/W, 73\% improvement in I/s/W/mm$^2$, and 10$\times$ improvement in I/s/W/footprint in \df{} over WS in 2D.}
As future work, we plan to improve other dataflows, such as output stationary, to utilize MIVs in \tech{} systolic arrays, 
\red{and include the fabrication cost for a comprehensive comparison.}

\section{Discussion}
\label{sec:discussion}
\rd{In this work, \df{} enables high bandwidth to minimize latency and improve energy efficiency. A 2D chip could conceivably support such a  high bandwidth but factors, such as routing congestion and fixed package area make such a chip design impractical. 
\bl{A 32 MB RRAM eliminates off-chip DRAM accesses for re-fetching weights in \resc{}, the largest DNN among those investigated, during its execution. DRAM accesses for other DNNs are also eliminated during their execution since their memory footprint is lower than that of \resc{}.
Similarly, 2 MB IFMAP and 2 MB OFMAP SRAMs eliminate the need for re-fetching IFMAP/OFMAP during a \conv{} layer of a DNN investigated in this work. We calculate the SRAM requirement by adding the IFMAP and OFMAP sizes in a \conv{} layer obtained from its topology file.
Finally, to minimize the area difference across tiers, we select a 6-tier \tech{} architecture, in which the first tier has a 256x256 PE array, the second tier comprises the SRAMs, while the remaining tiers constitute a 32 MB RRAM distributed across 4 tiers.}
In case the weights do not fit on-chip, data movement between the on-chip memory and DRAM for the corresponding convolutional layer needs to be included, and will \bl{increase the} execution cycles and DRAM energy. In this case, if the DRAM accesses are very frequent, the system energy is likely to be dominated by DRAM energy \cite{horowitz20141}, which can reduce the benefits coming from \df{}.}

\rd{DNN and \tech{} systolic arrays co-optimization needs new research efforts as well because there exist interesting tradeoffs between accuracy and latency. For instance, \resc{} has a higher top-1 accuracy on ImageNet (78.2\%) than \mnet{} (74\%) \cite{Guo_2020_CVPR}. However, in the systolic array configuration considered in our work, \mnet{} may be preferred due to the long latency of \resc{} in 2D. On the other hand, \df{} results in comparable latencies of the two DNNs and, as a result, \resc{} leads to better co-optimization of latency and accuracy.} \bl{Thus, \df{} can be plugged into a design optimization framework to co-optimize DNNs and \tech{} systolic arrays.}

\bibliographystyle{IEEEtran}
\bibliography{main}
\end{document}